\documentstyle[preprint,pra,aps]{revtex}

\begin{document}
\draft
\tightenlines

\title{\bf Statistics of electromagnetic transitions as a signature of
chaos in many-electron atoms}
\author{V. V. Flambaum, A. A. Gribakina, and G. F. Gribakin\cite{emg}}
\address{School of Physics, University of New South Wales,
Sydney 2052, Australia}
\date{\today}
\maketitle

\begin{abstract}
Using a configuration interaction approach we study statistics of the
dipole matrix elements ($E1$ amplitudes) between the 14 lower states with
$J^\pi=4^-$ and 21st to 100th even states with $J=4$ in the Ce atom
(1120 lines). We show that the distribution of the matrix elements is
close to Gaussian, although the width of the Gaussian distribution, i.e.
the root-mean-square matrix element, changes with the excitation energy.
The corresponding line strengths are distributed according to the
Porter-Thomas law which describes statistics of transition strengths
between chaotic states in compound nuclei. We also show how to use
a statistical theory to calculate mean squared values of the matrix
elements or transition amplitudes between chaotic many-body states.
We draw some support for our conclusions from the
analysis of the 228 experimental line strengths in Ce [J. Opt. Soc. Am.
{\bf 8}, 1545 (1991)], although direct comparison with the calculations is
impeded by incompleteness of the experimental data. Nevertheless,
the statistics observed evidence that highly excited many-electron states
in atoms are indeed chaotic. 
\end{abstract}
\vspace{1cm}
\pacs{PACS: 31.10.+z, 32.70.Cs, 31.50.+w, 05.45.+b\\
Submitted to Phys. Rev. A}
\maketitle

\section{Introduction}\label{intr}

The aim of this work is to present more evidence that excitation spectra
of complex open-shell atoms, and probably any other atom at sufficient
excitation energies, display clear quantum chaotic features. This
phenomenon is caused by strong mixing of many-electron excited states by
the residual two-body Coulomb interaction. It manifests in particular in
a Gaussian statistics of the $E1$ amplitudes for these states.

Since the time of Bohr's hydrogen atom theory atoms were considered as
perfectly regular dynamical systems. As the classical
theory of chaos evolved it became apparent that highly excited atomic
states in the Rydberg range could become chaotic if an external field is
applied \cite{atomsSC}, as long as the underlying classical motion is
chaotic.

On the other hand, it was also due to Bohr that the notion of compound
nuclei was introduced in physics. The behaviour of these highly excited
nuclear states is essentially quantum-mechanical. Nevertheless, they
display a number of chaotic properties. For example, the statistics of
their energy spectra show certain universal features, and transition
amplitudes involving compound states obey Gaussian statistics \cite{BM}.
To describe these properties it was suggested by Wigner that the
Hamiltonian of a compound nucleus could be modeled by a random matrix,
and different characteristics found by averaging over ensembles of
such matrices (see reviews \cite{Brody,Guhr}).

The first insight into quantum chaotic properties of complex atoms was
given by Rosenzweig and Porter \cite{RP} who analyzed experimental
spectra of some neutral atoms and showed that in heavy open-shell atoms
the spectral statistics are similar to those of compound nuclei. That
analysis was later extended and refined in \cite{Camarda}. Of course, the
study of eigenvalues provides valuable information about the
system. On the other hand, the spectral statistics observed in heavy
open-shell atoms are similar to those of the hydrogen atom in a strong
magnetic field \cite{Del}, or even a particle in a 2-dimensional
classically ergodic billiard \cite{billiard}. However, the eigenstates
of these quantum systems must be completely different, and it is clear
that the eigenvalue statistics cannot really tell us much about the
origin of chaotic behaviour, or indeed the structure of the chaotic
eigenstates.

The first inquiry into the possibility of chaos in the eigenstates of
complex atoms was done by B. Chirikov \cite{Chir}. He studied configuration
compositions of eigenstates of the Ce atom using data from the tables
\cite{martin}, and came to the conclusion that the `eigenfunctions are
random superpositions of some few basic states'. Inspired by that work
we conducted an extensive numerical study of the spectra and eigenstates
of complex open-shell atoms, using the rare-earth atom of Ce as an example
\cite{Ce,band,AIS}. This allowed us to investigate many-body
quantum chaos in a real system. We showed that atomic excited states are
in fact similar to nuclear compound states and developed a statistical
approach for analyzing their properties.

Unlike eigenvalues, the eigenfunctions are not observable directly.
To probe the structure of the chaotic eigenstates one can look at the
transition probabilities or matrix elements of some external perturbation
coupling them to each other, or to regular, simple eigenstates (like the
ground state). The matrix elements involving chaotic eigenstates must have
Gaussian statistics. We showed that its main characteristics -- the
mean squared value of the matrix element between the chaotic multiparticle
states (compound states), can be calculated in terms of statistical
parameters of the eigenstates and single-particle amplitudes and
occupation numbers of the orbitals present in the compound states
\cite{Ce,FV}.

In this work we have chosen the quantity most easily accessible
experimentally -- the $E1$ amplitudes. It also gives us an opportunity
to look for experimental signatures of chaos in the Ce atom using the work
by Bisson {\em et al} \cite{experCe}, where over 200 
line strengths were measured for transitions between a large
number of levels within 3.5 eV of the ground state. It should be mentioned
that there are many other possible atomic systems to search for
quantum chaos, e.g., in doubly excited states and inner-shell excitation
spectra of alkaline-earth atoms \cite{Con1,Con2,Con3}, or even multiply
excited states of light atoms \cite{Vaeck}.

\subsection{Chaotic many-body states}

Let us now recall briefly what chaotic many-electron atomic eigenstates
are. Suppose one uses a basis of some single-electron orbitals (e.g.,
the Hartree-Fock ones) to construct many-electron basis states
$|\Phi _k\rangle $. The states $|\Phi_k\rangle$ can be taken as
single-determinant states corresponding to certain configurations of a few
valence electrons, or constructed from them through some coupling scheme
to be of definite total angular momentum $J$. The true atomic eigenstates
\begin{equation}\label{A}
|\Psi_ i \rangle =\sum _kC_k^{(i)}|\Phi _k\rangle ~,\qquad
\Bigl( \sum _kC_k^{(i)2}=1\Bigr) ~,
\end{equation}
and eigenvalues $E^{(i)}$ are obtained by diagonalizing the Hamiltonian
matrix  $H_{jk}\equiv \langle \Phi _j|\hat H| \Phi _k\rangle $. The
coefficients
$C_k^{(i)}$ describe mixing of the basis states by the residual Coulomb
interaction. In the multi-electron excitation range the number of basis
states $|\Phi _k\rangle $ formed by distributing several electrons among
a few open orbitals is large. Many of these states are nearly degenerate
and the mean spacing between the basis state energies $E_k\equiv H_{kk}$ is
likely to be smaller than the typical value of the off-diagonal matrix
element $H_{jk}$. In this situation the basis states are strongly mixed
together \cite{note}.

Apart from a few lowest levels, each of the eigenstates is a superposition
of a large number of basis states. Of course, by a simple perturbation
theory argument, the mixing must be weak for distant basis states (large
$|E_j-E_k|$). The strong mixing takes place within a certain energy range
$|E_j-E_k|\lesssim \Gamma =2\pi V^2/D$, where $D$ is the mean level
spacing, $V^2=\overline{H_{jk}^2}$, and $\Gamma $ is called the
{\em spreading width}, since it characterizes the spread of the eigenstates
to which a given basis state contributes noticeably. One can estimate the 
number of {\em principal components}, i.e. those that contribute
significantly to a given eigenstate (\ref{A}), as $N\sim\Gamma /D$.
The coefficients $C_k^{(i)}$ corresponding to the principal components
have typical values  $|C_k^{(i)}|\sim 1/\sqrt{N}$. Their statistics is
close to that of independent random variables, and tends towards Gaussian
when the mixing is strong. In this case even the single-electron orbital
occupancies are far from integer and only the total angular momentum,
parity and the energy itself remain good quantum numbers \cite{Ce}. Thus,
we can talk about
{\em quantum chaos} in the system. This situation is similar to that in
compound nuclei and the corresponding chaotic eigenstates can be called
atomic compound states.
The model configuration interaction calculations performed for Ce
produced a value of $\Gamma \sim 2$ eV, and demonstrated the
existence of a dense spectrum of chaotic compound excited states
with $N\gtrsim 100$ ($D\sim 0.01$ eV) just few eV from the ground
state \cite{Ce}.

\section{Matrix elements between chaotic states}\label{mael}

Consider two chaotic many-body states (compound states, for
short) that are superpositions of large numbers of basis states,
$|\Psi _1\rangle =\sum _{k}C_k^{(1)}|\Phi_k\rangle $ and
$|\Psi _2\rangle =\sum _{j}C_j^{(2)}|\Phi_j\rangle $.
If the expansion coefficients $C_k^{(i)}$ are random, the matrix element
of some operator $\hat M$
\begin{equation}\label{m21}
\langle \Psi_2|\hat M|\Psi_1\rangle =\sum_{jk}C_j^{(2)}
\langle \Phi_j|\hat M|\Phi_k\rangle C_k^{(1)}
\end{equation}
is a sum of a large number of almost uncorrelated random items
\cite{FGI96}. Therefore, one should expect that such matrix elements
display Gaussian statistics with zero mean. Hence, the probability
distribution of the matrix elements between compound states can be
characterized by their mean squared value alone.

If $\hat M$ is a single-particle operator, e.g., the electric
dipole moment $\hat D=\sum _{\alpha \beta }
\langle \alpha |d|\beta \rangle a_\alpha ^\dagger a_\beta $
($\alpha $ and $\beta $ are single-particle states), it is
convenient to express its matrix elements in terms of the matrix
elements of the density matrix operator $\hat \rho _{\alpha \beta }
=a_\alpha ^\dagger a_\beta $,
\begin{equation}\label{Mel}
\langle \Psi _2|\hat D|\Psi _1\rangle =\sum _{\alpha \beta }\langle 
\alpha |d|\beta \rangle \langle \Psi _2|a_\alpha ^\dagger a_\beta 
|\Psi _1\rangle =\sum _{\alpha \beta}d_{\alpha \beta}\rho^{(21)}_
{\alpha \beta }~,
\end{equation}
where $\rho^{(21)}_{\alpha \beta }\equiv \langle \Psi _2|\hat \rho _
{\alpha \beta}|\Psi _1 \rangle $.

In \cite{FV} and \cite{Ce} a statistical approach to calculation of
mean squared matrix elements between compound states has been
developed. It is first based on the assumption that contributions from
different single-particle transitions $\beta \rightarrow \alpha $ in the
matrix element (\ref{Mel}) are uncorrelated. The mean squared value is then
given by
\begin{equation}\label{M2}
\overline{ |\langle \Psi _2|\hat D|\Psi _1\rangle |^2}=
\sum _{\alpha \beta}|d_{\alpha \beta}|^2
\overline{ |\rho^{(21)}_ {\alpha \beta }|^2}~,
\end{equation}
where averaging is done over a number of compound states around $\Psi _1$
and/or $\Psi _2$. The mean squared value of the density
matrix operator $\overline{ |\rho^{(21)}_ {\alpha \beta }|^2}$ is
expressed in terms of the parameters of the compound states 1 and 2
(i.e. their energies and spreading widths), and the average occupation
numbers of the single-particle states $\alpha $ and $\beta $.

In a spherically symmetric system where the states 1 and 2
are characterized by their total angular momenta $J_{1,2}$ and projections
$M_{1,2}$, the Wigner-Eckhart theorem applies, and it is convenient to deal
with the reduced matrix elements
$\langle \Psi _2\|\hat D\|\Psi _1\rangle $ independent of the projections
$M_{1,2}$. For example, the mean squared value of the zero-rank reduced
density matrix operator ($J_1=J_2\equiv J$ then) is obtained in the
following two forms \cite{Ce}: 
\begin{eqnarray}\label{answ0}
\overline {\Bigl| \rho ^{(21)0}_{nlj,n'l'j}\Bigr| ^2}=\cases {
D_1~\tilde \delta (\Gamma _1,\Gamma _2,\Delta )
\left( \frac {2J+1}{2j+1}\right) \langle n_{nlj}
(1-\frac{n_{n'l'j}}{2j+1})\rangle _2~,\cr
D_2~\tilde \delta (\Gamma _1,\Gamma _2,\Delta )
\left( \frac {2J+1}{2j+1}\right) \langle n_{n'l'j}
(1-\frac{n_{nlj}}{2j+1})\rangle _1~,\cr }
\end{eqnarray}
where $D_{1,2}$ are the mean level spacings near the states 1 and 2,
$n_{nlj}$ and $n_{n'l'j}$ are the orbital occupation numbers, and
$\tilde \delta $ is a ``finite-width $\delta $ function''. It depends
on the spreading widths $\Gamma_{1,2}$ of the compound states and
on the energy difference $\Delta =\omega _{n'l'j,nlj}-E^{(1)}+E^{(2)}$
between the transition frequency for the compound many-electron
states $E^{(1)}-E^{(2)}$ and the frequency $\omega _{n'l'j,nlj}$ of the
single-particle transition between the orbitals $nlj$ and $n'l'j$.
The function $\tilde \delta $ has a maximum at $\Delta =0$ and describes
the energy conservation for the compound states. Its width is determined
by the spreading widths $\Gamma _{1,2}$. Note
that $\langle \dots \rangle _{1,2}$ in Eq. (\ref{answ0}) denote averaging
of the occupation-number factors over the compound states 1 or 2. Note also
that the exact form of the function
$\tilde \delta (\Gamma _1,\Gamma _2,\Delta )$ depends of the
spreading of the compound states over the basis components, i.e. on the
``shapes'' of the eigenstates. In the simplest approximation this spreading
is described by the Breit-Wigner formula (see numerical studies in
\cite{Ce}) and $\tilde \delta $ is also a Breit-Wigner profile
\begin{equation}\label{BW}
\tilde \delta (\Gamma _1,\Gamma _2,\Delta )=
\frac {1}{2\pi }~\frac {\Gamma _1+\Gamma _2}{\Delta ^2+(\Gamma _1+
\Gamma _2)^2/4}~.
\end{equation}

To calculate the mean squared value of the $E1$ amplitude we now need a
formula for the reduced density matrix operator of the first rank.
Starting from the definition \cite{Ce}
\begin{equation}\label{rank1}
\rho ^{(21)1}_{nlj,n'l'j'}=(-1)^{J_2-M_2}
\left( {J_2\atop -M_2} {1\atop q} {J_1\atop M_1}\right)^{-1}\sum_{mm'}
(-1)^{j-m}\left( {j\atop -m} {1\atop q} {j'\atop m'}\right)~\rho^{(21)}_
{nljm,n'l'j'm'}
\end{equation}
for $q=0$ (linear polarization along the quantization axis) and assuming
that transitions between different magnetic sublevels $m$ are uncorrelated
we can derive a formula for the mean square of (\ref{rank1}), and then
use it to obtain the mean-squared $E1$ amplitude,
\begin{eqnarray}
\overline{| \langle \Psi _2 \| \hat D \| \Psi _1 \rangle |^2}
&=&\frac{2J_1+1}{3}~D_2\sum_{nlj,n'l'j'}
| \langle nlj \| d\| n'l'j' \rangle |^2\nonumber \\
&\times & \tilde \delta(\Gamma_1,\Gamma_2,\Delta) \left\langle \frac
{n_{n'l'j'}}{2j'+1} \left( 1- \frac{n_{nlj}}
{2j +1} \right) \right\rangle _1~,\label{answE1}   
\end{eqnarray}
analogous to the lower formula in Eq. (\ref{answ0}), or an alternative
form with $D_1$ and $\left\langle \frac
{n_{nlj}}{2j+1} \left( 1- \frac{n_{n'l'j'}}
{2j' +1} \right) \right\rangle _2$ on the right-hand side. The factor
$1/3$ on the right hand side of Eq. (\ref{answE1}) is due to the fact
that there are three final-state momenta $J_2=J_1,~J_1\pm 1$
accessible from a given $J_1$ by means of a dipole transition.
In deriving this expression an additional assumption has been made
that the occupancies of the $nljm$ and $n'l'j'm'$ states are
statistically independent, and the states with different $m$ within
the same $nlj$ shell are equally populated. This supposition influences
only the ``emptiness'' factors $\left( 1-\frac{n_{nlj}}{2j+1}\right) $,
which are close to unity anyway when the number of
single-electron states available is much greater than the number of
active electrons.

The square of the reduced dipole matrix element
$S(2,1)=| \langle \Psi _2 \| \hat D \| \Psi _1 \rangle |^2$
is called the strength of the line $1\rightarrow 2$, so
Eq.~(\ref{answE1}) allows one to estimate {\em mean line strengths}
for transitions involving compound states.

It is interesting to note that the statistical theory expression
(\ref{answE1}) satisfies the dipole sum rule \cite{sobel} (in atomic
units),
\begin{equation}\label{DSR}
\frac{2}{3}\sum _{J_2,E^{(2)}}\frac{E^{(2)}-E^{(1)}}{2J_1+1}
| \langle \Psi _2 \| \hat D \| \Psi _1 \rangle |^2 \approx n~,
\end{equation}
where $n$ is the number of active valence electrons included in the
configuration space of the problem.
To obtain this result one should replace summation over the final states
$2$ with integration over $dE^{(2)}/D_2$, take into account that
$\int (E^{(2)}-E^{(1)})\tilde \delta (\Gamma_1,\Gamma_2,\Delta )dE^{(2)}
=\omega _{nlj,n'l'j'}$ [see Eq. (\ref{BW})], neglect the ``emptiness''
factor $\left( 1-\frac{n_{nlj}}{2j+1}\right) \approx 1$ and use
$\sum _{n'l'j'}\langle n_{n'l'j'}\rangle _1=n$, and rely on the
single-particle sum rules for the orbitals $n'l'j'$ occupied in the
initial state $\Psi _1$,
\begin{equation}\label{spDSR}
\frac{2}{3}\sum _{nlj}\frac{\omega _{nlj,n'l'j'}}{2j'+1}
| \langle nlj \| d\| n'l'j' \rangle |^2 \approx 1~.
\end{equation}

\section{Numerical results for the Ce atom}

\subsection{Energy levels}

Cerium, $Z=58$, is the second of the lanthanide atoms. Its
electronic structure consists of the Xe-like $1s^2\dots 5p^6$ core and
four valence electrons. The atomic ground state is described by the
$4f6s^25d$ configuration with $J^\pi =4^-$ \cite{martin}.

The origin of the extremely complex and dense excitation spectra of the
rare-earth atoms is the existence of several open orbitals near
the ground state, namely $4f$, $6s$, $5d$, and $6p$,
or, in relativistic notation,
$4f_{5/2}$, $4f_{7/2}$, $6s_{1/2}$, $5d_{3/2}$, $5d_{5/2}$,
$6p_{1/2}$, and $6p_{3/2}$. These make a total of $N_s=32$ single-electron
states. For Ce with $n=4$ valence electrons there are about
$(N_s)^{n}/n!\approx 4\times 10^4$ possible many-electron states
constructed of them. If we allow for the two possible parities, about ten
possible total angular momenta $J$, and $2J+1$ different projections
(another factor of ten), there will be still hundreds of energy levels
within a given $J^{\pi }$ manifold.

In the present work we perform relativistic configuration interaction
calculations in the Hartree-Fock-Dirac basis analogous to those in
\cite{Ce}. In that work we limited ourselves to just 7 nonrelativistic
configurations constructed of the $4f$, $6s$, $5d$, and $6p$ orbitals,
for both odd and even states, which produced 260 and 276 states with
$J^\pi =4^-$ and $4^+$, respectively. To make the results more realistic
we have extended the configuration basis set by 9 odd and 23 even
nonrelativistic configurations. Basically, the additional configurations
were obtained by exciting one of the four electrons of an ``old''
configuration into the next orbital, e.g., the lowest even $4f^26s^2$
configuration would produce $4f6s^25f$, $4f6s^27p$, $4f^26s7s$, and
$4f^26s6d$ configurations. To keep the size
of the configuration space reasonable we included only the configurations
with mean energies within about 10 eV from the Ce ground state. This
increased the total number of $4^-$ and $4^+$ states to 862 and 1433,
respectively. Note that $J=4$ states have been chosen because these
manifolds are among the most abundant.

As a result, the level density $\rho (E)=\sum _i\delta (E-E^{(i)})$ has
increased greatly above 4 eV and become closer to that observed
experimentally. Of course, to be meaningful the level density must be
averaged over some small energy interval to obtain a smooth function
rather than a set of spikes. An alternative procedure is to look at the
cumulative number of levels
\begin{equation}
N(E)=\int _{-\infty}^E\rho (E')dE'~,
\end{equation}
which we present in Fig. \ref{fdens} for $J^\pi=4^+$ states.
Each $N(E)$ plot is a staircase of steps of the unit height occurring
at successive excited state energies. The level density can be easily
estimated from the slope of the $N(E)$ plot. The experimental data for
the 132 even levels with $J=4$ known from \cite{martin} is shown by the
solid-line staircase, and the energies are given with respect to either
experimental, or calculated ground state energy. They can be compared
with the dashed line
that shows $N(E)$ for our earlier small-basis calculation \cite{Ce} (276
states), and the dotted line for the present calculation (1433 states).
The improvement is obvious, however, the agreement is not perfect.
We believe that the remaining disagreement is not due to some missing
configurations in the CI calculation, but rather due to an overall
``softening'' of the spectra due to screening of the Coulomb repulsion
between the valence electrons by the electrons of the core \cite{screen}.
In the CI language this effect is produced by the high-energy excitations
of the valence electrons into the continuum together with the electron
excitations from the core.

Two typical features can be observed in the spectra of complex atoms
\cite{Camarda}. The first clearly seen in Fig. \ref{fdens},
is the rapid increase of the level density $\rho (E)$ with
energy \cite{Ryd}. Its origin is purely combinatorial -- the larger the
excitation energy, the greater the number of ways it can be distributed
among a few single-particle excitations. In the independent-particle
model this dependence is described by the following exponent \cite{BM}:
\begin{equation}\label{rho}
\rho _a(E)=\rho_0 \exp \left( a \sqrt{E-E_g}\right) ,
\end{equation}
where $\rho_0$ and $a$ are some constants, and $E_g$ is the ground state
energy of the system. This dependence also follows from the
thermodynamic definition of the temperature,
$T^{-1}=d\{ \ln [\rho (E)]\} /dE$,
combined with the estimates of the average number of excited Fermi
particles, $n_{\rm ex}\propto T$, and that of the excitation  energy,
$E-E_g\sim n_{\rm ex}T$. The experimental spectra of rare-earth atoms and
their ions examined in \cite{Camarda} are in agreement with
Eq. (\ref{rho}).

Figure \ref{fdens} shows that the calculated cumulative level number plot
is fitted well by
\begin{equation}\label{NE}
N(E)=\int _{E_g}^E \rho _a(E')dE'
\end{equation}
with $\rho_0=0.65$ eV$^{-1}$, $a=2.55$ eV$^{1/2}$, and the ``ground state''
energy of the $4^+$ sequence $E_g$ shifted by 0.25 eV up from the true
$J^\pi=4^-$ ground state of Ce. Thus, Eq. (\ref{rho}) gives a good overall
fit of the calculated level density below 6 eV.

The second feature typical for the spectra of complex many-body systems is
level repulsion. It is a basic quantum mechanics fact that two levels
with identical quantum numbers cannot be degenerate if they are
coupled by a non-zero matrix elements -- they ``repel'' each other. In
quantum chaotic systems this repulsion is characterized by the Wigner
level spacing distribution
\begin{equation}\label{wigner}
P(s)=\frac{\pi s}{2}e^{-\pi s^2/4},
\end{equation}
where $s$ is the nearest-neighbour level spacing normalized so that
$\overline s=\int s P(s)ds =1$. Equation (\ref{wigner}) shows that
the probability of finding small level spacings is indeed vanishingly
small. As we pointed out in the Introduction, spectral statistics
do not tell much about the eigenstates of the system. However,
Eq. (\ref{wigner}) is still a good test for some possible hidden quantum
numbers, e.g., the total spin or orbital momentum,
which might characterize atomic eigenstates besides $J^\pi $. If these
do exist, small level spacings (``degeneracies'') will be more abundant
than predicted by Eq. (\ref{wigner}). These statistics were checked
for many experimental \cite{RP,Camarda,Ce,Con1,Con3} and calculated
\cite{Ce,Con2} complex atomic spectra, as well as for molecular
vibronic spectra \cite{molec}.

As seen from Fig. \ref{fdens} the level density changes significantly for
the first 500 levels of the calculated spectrum. To analyze the
distribution of the corresponding level spacings we use the analytical
density fit $\rho _a(E)$ to normalize the spacings:
\begin{equation}\label{spac}
s_n=(E_{n+1}-E_n)\rho _a(E_n).
\end{equation}
Their distribution shown on the inset in Fig. \ref{fdens} is in reasonable
agreement with the Wigner formula. The deviations are probably due to
the long-range fluctuations of the level density, not accounted for
by the simple exponential (\ref{rho}). In the previous calculation
\cite{Ce}, where only the lowest orbitals of each symmetry were included,
we also observed the Wigner distribution. When orbitals with higher
principal quantum numbers become involved (as seen from Fig. \ref{fdens}
above 3.5 eV) the spatial extent of the eigenstates increases. This should
cause a  decrease of the residual Coulomb interaction between the
electrons. On the other hand, the level spacings also become smaller. As a
result, the state mixing at these excitation energies remains strong,
which is confirmed by the agreement with the Wigner distribution,
and the eigenstates are chaotic. Our estimate of the number of principal
components $N$ shows that it becomes even greater as the energy increases,
in accord with the estimate $N\sim \Gamma /D\sim 300$ ($\Gamma \sim 1$ eV,
and the mean level spacing $D\approx 0.003$ eV at $E\approx 6$ eV).

\subsection{Dipole matrix elements}

In Sec. \ref{mael} we explained that matrix elements involving
chaotic compound states should have Gaussian statistics, and the mean
squared value of the matrix elements could be estimated in terms of
some average characteristics of the compound states. In this section
we concentrate on the dipole matrix elements ($E1$ amplitudes)
$d_{ik}=\langle \Psi _i^{4^+}\|\hat D\|\Psi _k^{4^-}\rangle $ between the
14 lowest states with $J^{\pi} = 4^-$ and 80 consecutive $4^+$ states
obtained numerically in our CI calculations of Ce. We have
chosen this energy region to cover the range explored in the experiment
\cite{experCe}, where absolute values were derived for 228 of the most
intense lines of neutral Ce between 10706 and 22184 cm$^{-1}$.

Of course, low-lying atomic states, e.g., the ground state, have
well-defined configuration composition and are not chaotic, hence, the
$E1$ amplitudes between them should not be distributed in any particular
statistical way. However, the matrix elements (\ref{m21}) will become
random (and close to Gaussian) as soon as at least one of the
states involved, the initial or the final, moves into the compound-state
energy range and becomes a superposition of many random components.
Besides that, the mean squared value of the matrix element is expected to
show some smooth secular variation with the energy of the states involved.
For these reasons we skip the first 20 states with $J^\pi =4^+$ and analyze
the statistics of the $14\times 80=1120$ $E1$ amplitudes for the
following 80 even states by grouping them in bunches of twenty: 21--40,
41--60, 61--80, and 81--100, which correspond to the mean excitation
energies of 2.49, 2.95, 3.40, and 3.70 eV above the atomic ground state
(the mean energy of the lowest 14 odd states is 0.68 eV). Thus, each plate
in Fig. \ref{fcomp} shows the
distribution of the 280 reduced dipole matrix elements together with their
root-mean-square (r.m.s.) value. Also shown in Fig. \ref{fcomp} are the
Gaussian distributions $g(d)=\exp (-d^2/2d_0^2)/\sqrt{2\pi d_0^2}$,
where the root-mean-square parameter $d_0$ has been adjusted to minimize
$\chi ^2$ around the center of the histogram. The values of $d_0$ and
$\chi ^2$ are given in Table \ref{comp}.

Two effects can be seen in Fig. \ref{fcomp}. First, the distributions of
the matrix elements are indeed close to Gaussian. Second, the width
of the distributions (the mean squared value of the matrix elements)
varies with the energy of the even states. It is mostly this effect that
is responsible for the visible discrepancies between the histograms and
the Gaussian fits. To eliminate it we can use a running average procedure
to normalize the amplitudes:
\begin{equation}\label{dnorm}
d_{ik}^{({\rm n})}\equiv \frac{d_{ik}}{\langle{d^2}\rangle _i^{1/2}}~,
\end{equation}
where $\langle{d^2}\rangle _i^{1/2}$ is the
r.m.s. value over the 14 odd states, calculated for every even state $i$.
Figure \ref{fnorm} confirms that the 1120 normalized $E1$ amplitudes for
the 21--100 even states are distributed according to the normal law.
The inset shows the dependence of the r.m.s. $E1$ amplitude
$\langle{d^2}\rangle _i^{1/2}$ on the energy of the even state
$E^{(i)}$. Fluctuations aside, it is in agreement
with the r.m.s. values calculated from the statistical theory,
Eq.~(\ref{answE1}), at the energies of the 30th, 50th, 70th and 90th even
states. The numerical values of the r.m.s. $E1$ amplitudes are listed
in Table \ref{comp}.

Note that we have chosen Eq. (\ref{answE1}) with 1 standing for the odd
states and 2 for the even ones. In our numerical example we consider the
dependence of the r.m.s. $E1$ amplitude on the energy of the even states,
and keep the odd states the same. Therefore, as in Eq.~(\ref{answE1}),
we only need to know the average occupation numbers for the lowest 14 odd
states, and the result depends on the final even state via its energy
$E^{(2)}$, mean level spacing $D_2$ and spreading width $\Gamma _2$. As we
saw in our previous calculations \cite{Ce}, the even states of Ce
with $J=4$ become very much chaotic at excitation energies of just 2 eV,
i.e. from the 20th level up. Also, as earlier in \cite{Ce}, we use average
configuration energies rather than single-particle Hartree-Fock energies
to determine the transition frequencies $\omega _{n'l'j,nlj} $ needed for
calculation of
$\Delta $ in Eq. (\ref{answE1}). The ground state of Ce is described as
$4f6s^25d$, however, the dominant configuration among the 14
lowest odd states is $4f6s5d^2$, and we used it to calculate the
transition energies. For example, the energy of the $6s-6p$ transition
$\omega _{6p,6s}$ was determined as the difference between the average
energies of the $4f5d^26p$ and $4f6s5d^2$ configurations. Physically,
this corresponds to choosing a particular mean field close to that of
the low-lying odd states of Ce for calculation of the transition energies.
It should be mentioned though that the results obtained with the
Hartree-Fock frequencies $\omega _{nlj,n'l'j'}= \epsilon _{nlj}-
\epsilon _{n'l'j'}$ were not too different.

Gaussian statistics of the dipole matrix elements result in the
Porter-Thomas (PT) distribution of the line strengths $S(i,k)=d_{ik}^2$
\begin{equation}\label{PT} 
f(S)=\frac{1}{\sqrt{2\pi S\overline{S}}}{\exp\left(-\frac{S}
{2\overline{S}}\right)},
\end{equation}
where $\overline{S}$ is the mean line strength. Divergence of this
function at small $S$ means that if the $E1$ amplitudes are Gaussian,
there should be many weak lines in the spectrum.
Earlier evidence of the PT statistics of line strengths
can be found in calculations of dipole excitations in complex atoms
\cite{Karazia}, and transitions between the vibronic levels in molecules
measured in \cite{molec}.

\section{Analysis of experimental data}

In\cite{experCe} absolute values of $gA$ were obtained for 228 of the most
intense observed lines between 10706 and 22184 cm$^{-1}$ in Ce. It is
interesting to analyze these data to see whether they support our
theoretical and numerical considerations.

The values of $gA$ listed in \cite{experCe} are defined as
$gA=(2J_k+1)A_{ki}$, where
\begin{equation}\label{Ak}
A_{ki}=\frac{4e^2\omega _{ki}^3}{3\hbar c^3(2J_k+1)}|
\langle i \|\hat D\|k\rangle |^2 
\end{equation}
is the $E1$ transition rate from the upper level $k$ into the lower level
$i$ \cite{sobel}. We use the experimental values of $gA$, $J_k$ and
transition frequencies $\omega _{ki}$ to extract values of the line
strengths
\begin{equation}\label{mel}
S(i,k)\equiv |\langle i\|\hat D\|k \rangle |^2 
=gA\frac{3\hbar c^3}{4e^2\omega_{ki}^3}~.
\end{equation}
In Fig. \ref{Port} the probability distribution of the 228 experimental
line strengths is shown. Compared to the expected PT formula 
(\ref{PT}), there is a clear lack of small line strengths. Nevertheless,
the decreasing part of the histogram can be fitted well by a PT
distribution with an additional normalization factor $A$,
\begin{equation}\label{PTA}
f_A(S)=\frac{A\exp (-S/2\overline{S})}{\sqrt{2\pi S\overline{S}}}~,
\end{equation}
shown in Fig. \ref{Port} by a solid line for $A=2.07$ and
$\overline{S}=3.3$ a.u. that minimize $\chi ^2$ for the 22 bins with
$S>3$~a.u.

It would be tempting to say that the excellent agreement
between the PT curve and the histogram is a confirmation of the
Gaussian statistics of the $E1$ amplitudes in Ce. The value of $A$ would
then indicate that about one half of all lines are missing in the
experimental data. However, the value of $\overline{S}=3.3$ a.u.
corresponds to the r.m.s. $E1$ amplitude of 1.8, which is more than 2 times
greater than our numerical results in Fig. \ref{fcomp} and Fig. \ref{fnorm}
(inset)
and in Table \ref{comp}. On the other hand, the experimentally observed
228 lines include transitions between levels with various total angular
momenta between $J=1$ and 8 ($|J_i-J_k|\leq 1$, of course)
whereas we have about 500 hundred lines with just $J_i=J_k=4$ in our
calculation in the analogous energy range. This means that in
\cite{experCe} only the strongest 10\% or less of all lines have in fact
been measured. The very suggestive agreement with the PT distribution in
Fig. \ref{Port} should then be considered as merely fortuitous.

It is worth noting that in experiment the lines are selected by their
intensities proportional to $gA$, rather then strengths. Hence, even lines
with large strengths can be omitted if their frequencies are small.
Let us look at the simplest model of this effect and see how it
influences the observed strengths distribution. Assume that transitions
in a certain frequency range $0<\omega <\omega _{\rm max}$ are studied,
and different values within this interval are equally probable. The
observed intensities of the lines are proportional to $\omega ^3S$. If we
assume that there is a minimal threshold intensity that can be registered,
the original PT distribution of strengths would be modified as follows:
\begin{eqnarray}\label{MPT}
f_1(S)=\cases {
0~,\quad S\leq S_0\cr
\frac{A}{\sqrt{2\pi S \overline{S}}}\exp \left( -\frac{S}{2\overline{S}}
\right)
\left[ 1-\left( \frac{S_0}{S} \right) ^{1/3}\right] ~,\quad S>S_0\cr }
\end{eqnarray}
where $S_0$ is the minimal strength that can be observed at
$\omega =\omega _{\rm max}$, and $A$ is the normalization factor.
As seen from Fig. \ref{Port}, Eq. (\ref{MPT}) also gives a very good fit
of the experimental data with $\overline{S}=2.4$, $S_0=1.12$ and $A=6.22$
\cite{note1}. Note, however, that the new value of the mean line strength
is 1.5 times smaller than the one we had from the pure PT fit. Therefore,
the assumptions used in our processing of the experimental data affect
the estimates of the experimental r.m.s. $E1$ amplitudes, and we should
not be too concerned about the apparent disagreement with our numerical
calculations. Besides that, extraction of absolute line strengths from
the experimental data is not free from uncertainties estimated in
\cite{experCe} at 10--20\%.

For 30 transitions in Ce the $gA$ values were obtained more accurately
from branching ratios and delayed photoionization measurements of
lifetimes (\cite{experCe}, Table 2). When we look at the statistics of
the corresponding line strengths (Fig. \ref{Port}, inset) and compare it
with the
PT distribution (\ref{PT}), a value of $\overline{S}=2.15$ is obtained,
much smaller than the estimates of $\overline{S}$ from the statistics
of the 228 lines. Thus, it appears that to make firm conclusions about
Gaussian statistics of the $E1$ amplitudes a much more thorough
experimental survey is needed. On the other hand, even relative
measurements of a large number of line strengths could be very valuable
for examining these statistics \cite{molec}.

\section{ Conclusions}

In this work we have extended the configuration interaction approach
of \cite{Ce} to calculate large numbers of eigenstates in Ce. In agreement
with our earlier studies the energy level statistics indicate that the
simple configurational basis states are strongly mixed together by the
residual
electron interaction, and the only good quantum numbers in the spectrum
are parity and the total angular momentum. The total orbital momentum $L$
and spin $S$ are not conserved due to the spin-orbit interaction, whose
effect is dynamically enhanced, just as that of any other perturbation in
a chaotic many-body system \cite{Ce}.

The strong configuration mixing makes multielectron atomic eigenstates
chaotic. This in turn results in a Gaussian statistics of the matrix
elements for chaotic atomic eigenstates (compound states). This
understanding is fully
confirmed by our numerical calculations of the 1120 $E1$ amplitudes
between the 14 lowest $J^\pi =4^-$ states and 80 $J^\pi =4^+$ states above
2 eV. It is important that the parameter of the Gaussian, the
r.m.s. $E1$ amplitude, varies slowly with the excitation energy.
This effect should be taken into account when analyzing the statistics
of the matrix elements.

We also show that a statistical theory can be used to estimate mean
squared matrix elements involving compound states. It enables one to
express the answer in terms of the single-particle matrix elements and
occupation numbers, and parameters of the compound states, namely the
number of principal components and the spreading width. This approach
has already been applied to calculation of matrix elements between
compound states in nuclei \cite{FV}. It could be useful in various other
many-body systems, e.g., atomic clusters or quantum dots, where direct
diagonalization of the Hamiltonian matrix is not feasible because of
a huge size of the Hilbert space of the problem.

An attempt has been made to analyze existing experimental data for the
line strengths in Ce \cite{experCe}. It appears that the statistics
of the measured line strengths is compatible with the Porter-Thomas
distribution, with allowance for the missing weak lines. However, the
discrepancy between the calculated r.m.s. $E1$ amplitudes and those
inferred from the experimental data does not allow us to say that
the existence of quantum chaos in the Ce eigenstates has been confirmed
experimentally. To make this statement one would have to do a much more
complete survey and statistical analysis of the line strengths in the Ce
spectrum.

On the other hand, this means that a comparison between the experimental
and theoretical line strengths in Ce is not yet possible, even at the
level of their mean values. Theoretically, to calculate precisely the
dipole matrix elements between particular levels in the compound-state
energy range of complex atoms like Ce looks a prohibitively difficult
problem. Experimentally, identification of specific lines in enormously
complicated spectra is also a very difficult task. However, we would like
to suggest that extraction of mean characteristics from the experiment and
comparison with the corresponding theoretical estimates is a meaningful
way of exploring such complex systems. As a result, one might hope to
get a deeper insight into the existence of quantum chaos in many-body
systems on the whole, and in complex open-shell atoms, in particular.

\acknowledgments

We would like to thank O. P. Sushkov for useful discussions, and
acknowledge support of this work by the Australian Research Council.



\begin{table}
\caption{Root-mean-square $E1$ amplitudes for transitions between
the 14 $J^\pi =4^-$ and 80 $J^\pi =4^+$ states in Ce.}
\label{comp}
\begin{tabular}{cccccc}
 & \multicolumn{3}{c}{r.m.s. $E1$ amplitudes (a.u.)} & & \\
\cline{2-4}
even levels & $\left(\overline{d_{ik}^2}\right) ^{1/2}$\tablenotemark[1]
& from Eq. (\ref{answE1})\tablenotemark[2] & $d_0$\tablenotemark[3]
& $n$\tablenotemark[4]
& $\chi^2(n-1)$\\
\hline
21--40  & 0.853  & 0.813 & 0.729  & 21 & 24.4\\
41--60  & 0.891  & 0.746 & 0.824  & 21 & 23.9 \\
61--80  & 0.736  & 0.674 & 0.671  & 17 & 36.6 \\
81--100 & 0.627  & 0.566 & 0.632  & 17 & 24.7

\end{tabular}
\tablenotetext[1]{Obtained directly from the CI calculation.}
\tablenotetext[2]{Calculated from the statistical theory,
Sec.~\ref{mael}.}
\tablenotetext[3]{Values that minimize $\chi ^2$ for the Gaussian
fits shown in Fig. \ref{fcomp}.}
\tablenotetext[4]{Number of bins around the centre of the histogram
used for calculation of $\chi ^2$.}

\end{table}


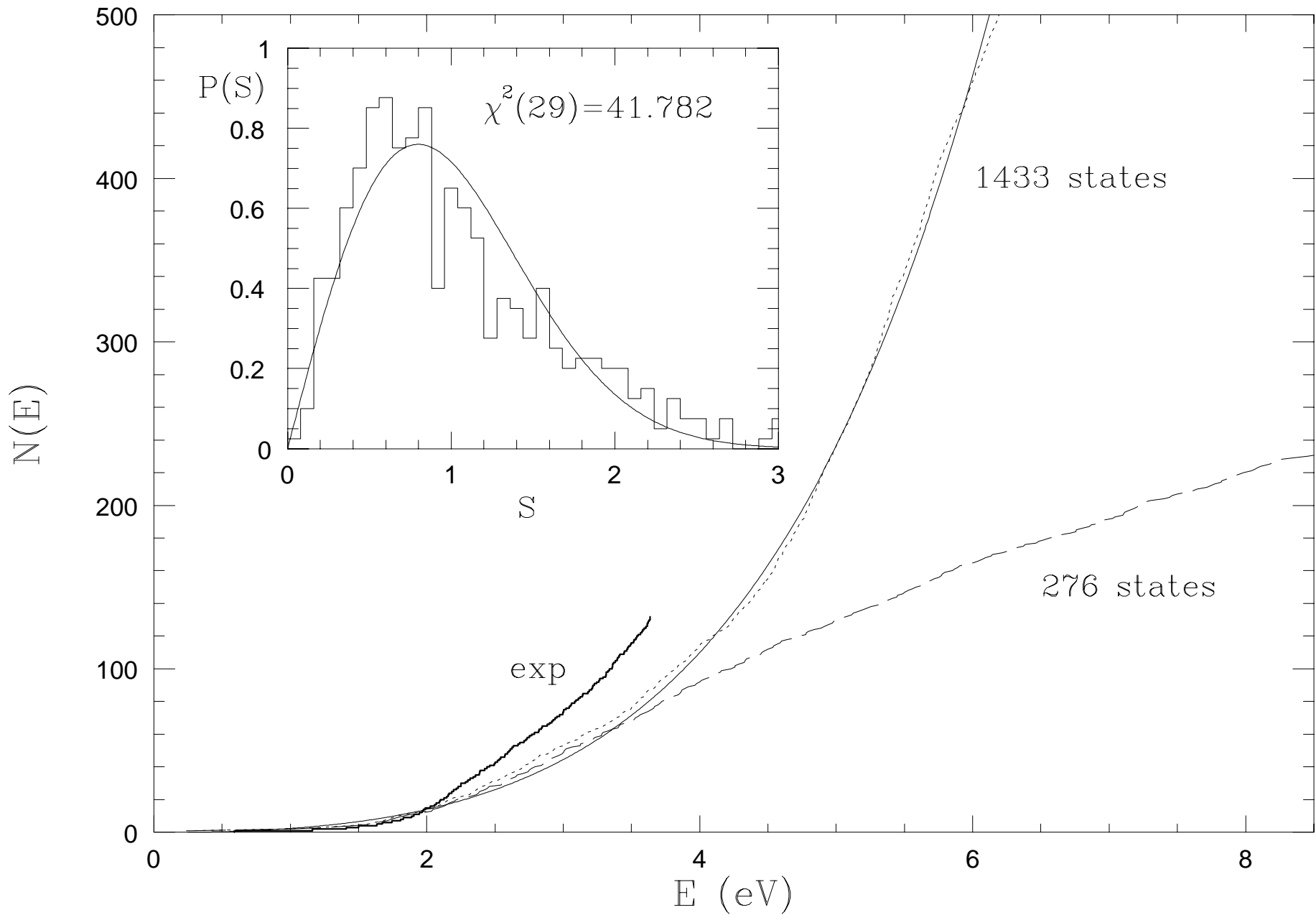
\begin{figure}
\caption{Energy spectra and level statistics of the $J^{\pi}=4^+$ states
in Ce. Dashed line shows the cumulative number of states $N(E)$ for the
calculation with 276 basis states {\protect \cite{Ce}}; dotted line is
the present calculation with 1433 basis states; thick solid line is $N(E)$
for 132 experimental levels from {\protect \cite{martin}}. Thin solid line
is the cumulative level number corresponding to the independent-particle
fit (\protect\ref{NE}). Shown on the inset are the statistics of the
normalized level spacings $s$ for the lowest 500 levels, compared with the
Wigner distribution (\protect\ref{wigner}). \label{fdens}}
\end{figure} 
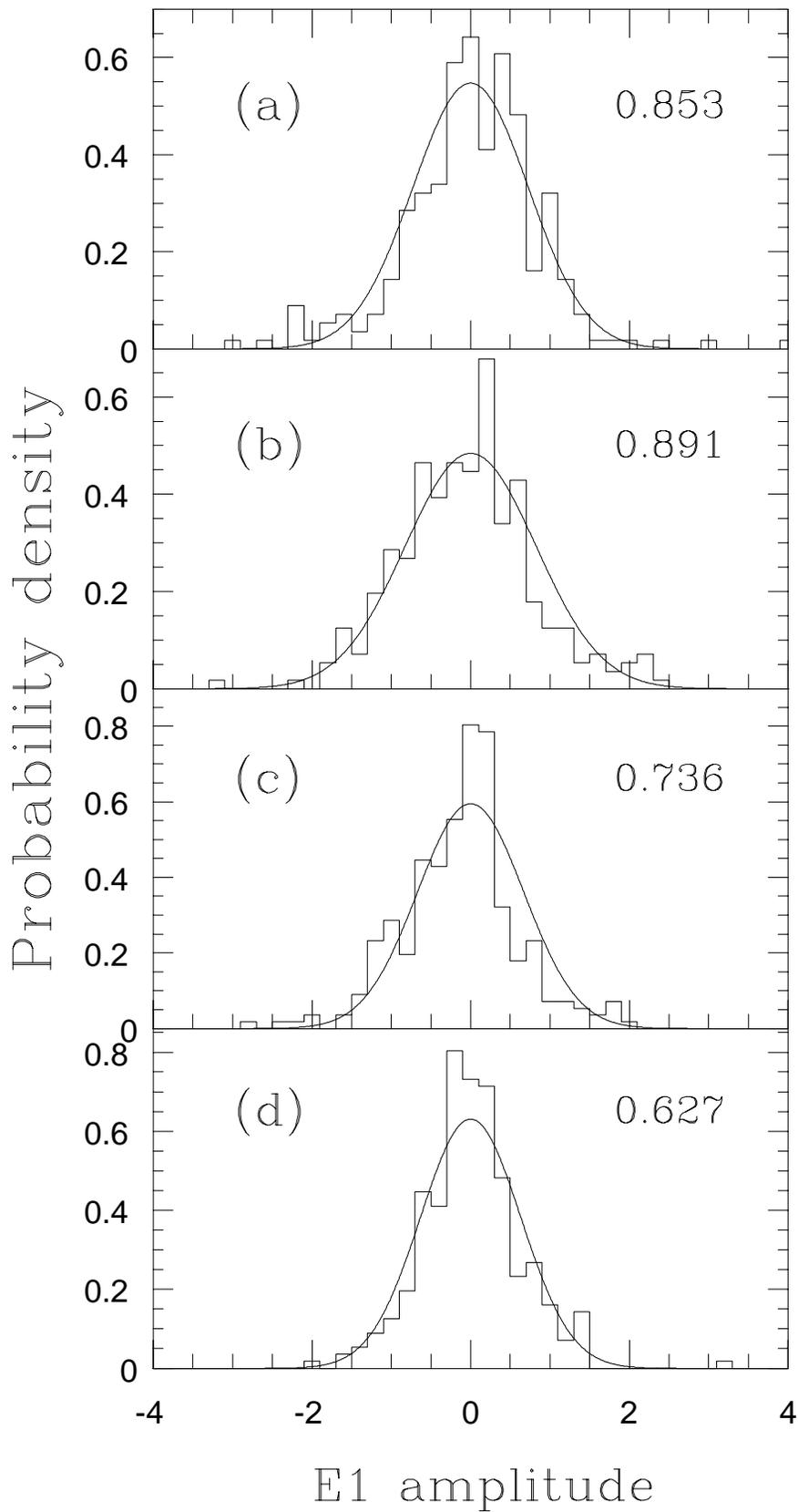
\begin{figure}[h]
\caption{Probability distributions of the $E1$ amplitudes in Ce for
transitions between the 14 lowest $4^-$ states and groups of 20 states
with $J^\pi =4^+$: (a) 21--40, (b) 41--60, (c) 61--80, and (d) 81--100.
R.m.s. values of the amplitudes are shown next to the histograms.
Smooth curves are Gaussian fits that minimize $\chi ^2$
for 21 to 17 central bins of the histograms (see table
\protect\ref{comp}).\label{fcomp}}
\end{figure}
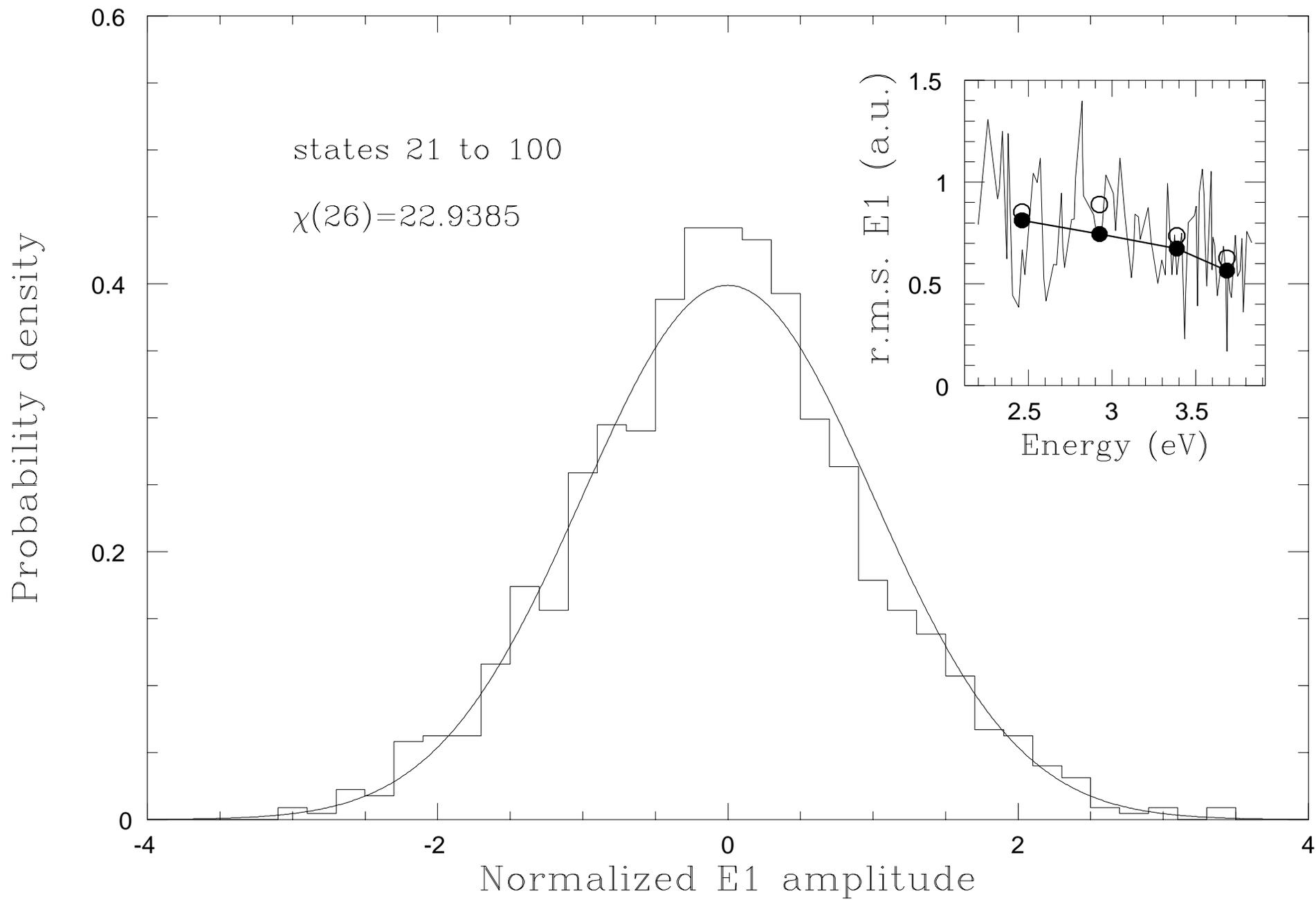
\begin{figure}[h]
\caption{Probability distributions of the normalized $E1$ amplitudes in
Ce for transitions between the 14 lowest $4^-$ states and 21--100 states
with $J^\pi =4^+$, compared to the normal distribution (solid line).
The inset shows the dependence of the r.m.s. $E1$ amplitude averaged over
the 14 odd states on the energy of the even state
(thin solid line). Solid circles connected by a thick solid line are the
r.m.s. values of the $E1$ amplitude obtained from the statistical theory,
Eq.~(\protect\ref{answE1}), at the energies of the 30th, 50th, 70th, and
90th even states, and open circles are values from the CI calculation
(see table \protect\ref{comp}).\label{fnorm}}
\end{figure}
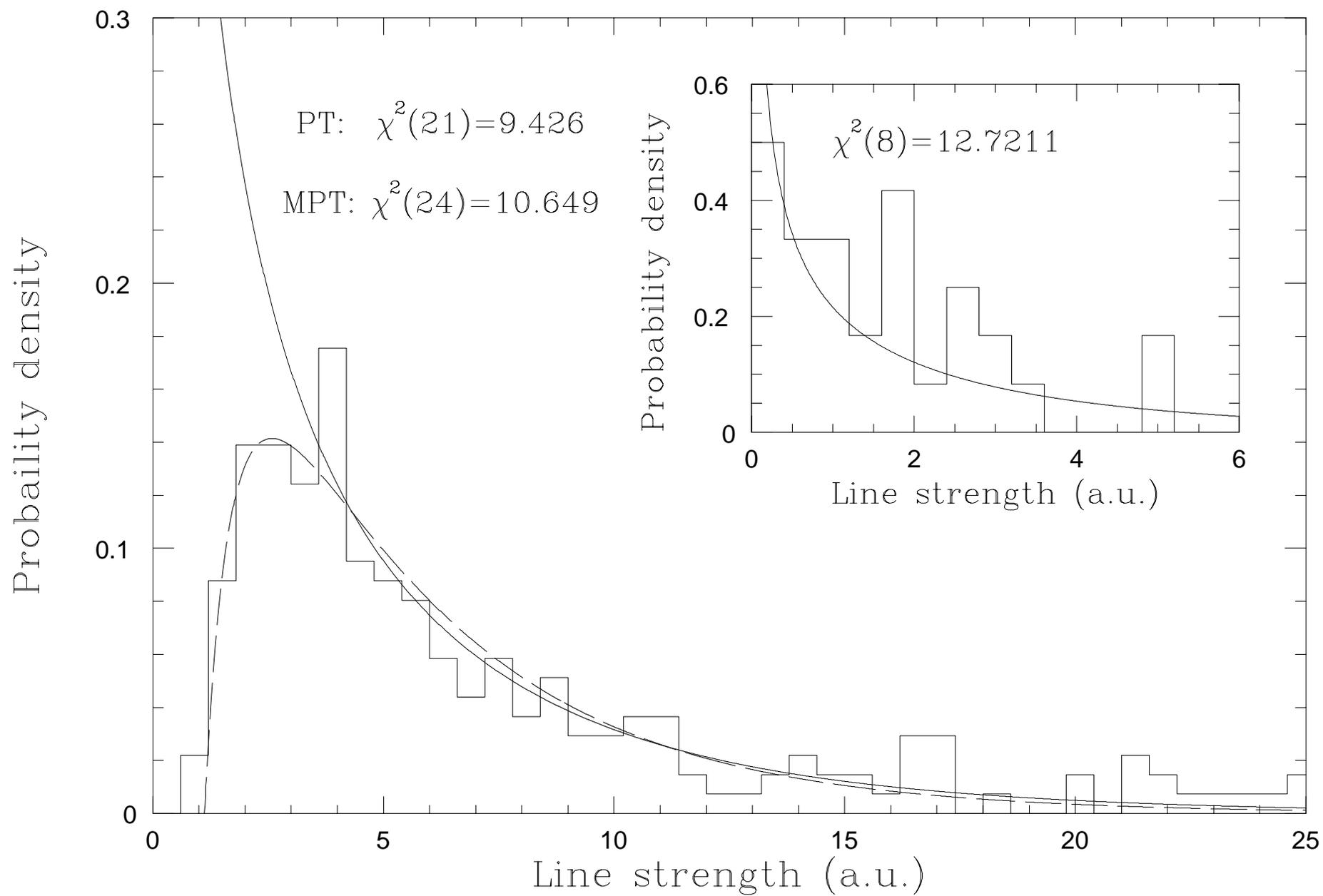
\begin{figure}
\caption{Comparison of the line strengths measured in Ce by Bisson
{\em et al} \protect \cite{experCe} with the Porter-Thomas and modified
Porter-Thomas distributions. Solid line is the PT distribution
(\protect \ref{PTA})
with $A=2.07$ and $\overline{S}=3.3$ a.u., and dashed line is the modified
PT distribution (\protect \ref{MPT}) with $A=6.22$, $S_0=1.12$ and
$\overline{S}=2.4$ a.u. Shown on the inset is the probability distribution
of the 30 lines measured from branching ratios and delayed photoionization,
fitted by a PT distribution with $\overline{S}=2.15$ a.u.
\label{Port}}
\end{figure} 


\begin{thebibliography}{10}
\bibitem{note1}The normalization integral $\int f_1(S)dS$ for this set of
values is equal to 0.902, as this fit, just as the pure PT fit, evidently
fails to describe the large number of line strengths greater than
15 a.u. observed in \cite{experCe}.
\end{thebibliography}
\end{document}